\documentclass[aps,prd,showpacs,amsmath,amssymb]{revtex4}
\usepackage{epsfig}

\begin{document}

\title{Abelian representation for nonabelian Wilson loops \\and  the Non -  Abelian Stokes theorem
on the lattice.}

\author{M.A.Zubkov}
 \email{zubkov@itep.ru}
\affiliation{ITEP, B.Cheremushkinskaya 25, Moscow, 117259, Russia}

\date{July, 12, 2003}

\begin{abstract}
We derive the Abelian - like expression for the lattice $SU(N)$ Wilson loop in
arbitrary irreducible representation. The continuum Abelian representation of
the $SU(N)$ Wilson loop (for the loop without selfintersections) that has been
obtained by Diakonov and Petrov appears to be a continuum limit of this
expression. We also obtain the lattice variant of a non -  Abelian Stokes
theorem and present the explicit expression for the matrix $\cal H$ used in the
Diakonov - Petrov approach.
\end{abstract}


\maketitle

\section{Introduction}

In order to visualize  multidimensional space the projection onto the lower
dimensional spaces can be used. The analogous idea is used in the non - Abelian
gauge theory, where the dynamics is described in terms of certain Abelian
variables. This method is called the Abelian Projection \cite{tH}.

One of the most important quantities we encounter in gauge theory is the Wilson
loop
\begin{equation}
W_q[{\cal C}] = \chi_q ({\rm P} {\rm exp} \,(i \int_{\cal C} A_{\mu} dx_{\mu}))
\end{equation}
Here $A$ is the gauge field (that belongs to the Lie algebra of the
correspondent gauge group), $\cal C$ is the closed contour, $\rm P$ denotes the
usual path ordering, and $\chi_q$ is the character considered in the
representation $q$ of the gauge group. For the application of the Abelian
projection method it is important to have an Abelian - like representation for
the Wilson loop. The representation of such kind was derived more than ten
years ago by D. Diakonov and V. Petrov \cite{DP}. For the case of the
irreducible representation $q$ of $SU(N)$ it has the form:
\begin{equation}
W_q[{\cal C}] = \int D \mu_{{\cal C},q}(g) {\rm exp} \,(i \int_{\cal C} {\rm
Tr} \, A_{\mu}^g\, {\cal H}^q dx_{\mu}),\label{Wdp}
\end{equation}
where $g$ is the $SU(N)$ gauge transformation and $A_{\mu}^g = g A_{\mu} g^+ -
i g
\partial_{\mu} g^+$. Here ${\cal H}^q = \sum_{i = 1, ..., N-1} m_i H_i$, where
each $H_i$ denotes the basis element of the Cartan subalgebra of $su(N)$ and
$(m_i)$ is the
 highest weight of the representation. The matrices $H_i$ are
normalized in such a way that ${\rm  Tr}\, H_i H_j = \delta_{ij}$.  The
 gauge transformations are defined on the contour $\cal C$. The
measure $\mu_{{\cal C},q}(g)$ on the space of gauge transformations is
constructed using the invariant measure on the $SU(N)$ group. This construction
will be discussed in detail later.

Both in the monopole and in the $P$ - vortex confinement pictures
 the confining forces are induced by the magnetic flux contained
inside the contour that is correspondent to the worldline of the charged
particle \cite{P}. In the Abelian theory a Stokes theorem is used for counting
the magnetic flux. For counting the "magnetic flux" contained inside the
contour correspondent to the Wilson loop in a non - Abelian theory D. Djakonov
and V. Petrov suggested to use representation (\ref{Wdp}) in the following
form:
\begin{equation}
W_q[{\cal C}] = \int D \mu_{{\cal C},q}(g) {\rm exp} \,(i \int_{{\cal M}(\cal
C)} {\rm  Tr} \, \partial_{[\mu}A_{\nu]}^g\, {\cal H}^q d x_{\mu}\wedge d
x_{\nu}),\label{Ws}
\end{equation}
where $g$ is defined on the surface ${\cal M}(\cal C)$ that spans the contour
$\cal C$. The integral of a two - form over this surface is denoted by
$\int_{{\cal M}(\cal C)} ... d x_{\mu} \wedge d x_{\nu}$. The expression
(\ref{Ws}) is known as the non - Abelian Stokes theorem.

In the present work we consider a lattice variant of  the Abelian
representation for the non -  Abelian Wilson loop and the non -  Abelian Stokes
theorem on the lattice. The representation (\ref{Wdp}) was originally derived
in a continuum theory. In the later publications (see, for example,
\cite{DP1,KT}) the approximated lattice variant of (\ref{Wdp}) was derived.

It appears that one of the most fruitful methods of the investigation of the
 quantum field theory is the lattice approach. This approach
gives the exact meaning to any field correlator. It is possible both to define
and to calculate it using the lattice regularization. Without using a lattice
 numerical investigation this could become possible  only for
the Integrable Models.

Thus it seems important to derive the lattice variant of (\ref{Wdp}), which is
valid within the lattice $SU(N)$ theory {\it precisely} (i.e. not only in its
naive continuum limit). In this paper such a representation is derived. If we
assume that the gauge field is smooth enough, then the continuum limit of the
derived representation coincides with (\ref{Wdp}). The non - Abelian Stokes
theorem on the lattice commonly follows from the lattice Abelian representation
of the Wilson loop.

\section{Irreducible representations of the $SU(N)$ group}

In this section we briefly summarize the basic notations of group
representation theory that will be used in the next sections.

1. The space $\cal V$ of any irreducible representation of $SU(N)$ consists of
the tensors $\Psi_{i_1 i_2 ... i_r}$. The symmetry of $\Psi$ is defined by the
set of integer numbers $q_i \, (i = 1, ..., N-1)$ ($\sum_i q_i = r, \, q_i\geq
0 $), which is called the signature of the representation. The elements of
$\cal V$
 are the tensors
\begin{equation}
\Psi_{i^1_1 ...i^1_{q_1} ...i^{c_1}_1 ... i^{c_1}_{q_{c_1}}}
\end{equation}
For any $k \in \{1, ..., q_1\}$  the number of indices $i^k_l$ is denoted by
$c_k$. Let us arrange these indices as follows:
\begin{equation}
\begin{array}{ccccc}
i^1_1 & ... & i^1_i & ... & i^1_{q_1} \\
i^2_1 & ... & i^2_{q_2} \\
... \\
i^{c_1}_1 & ...
\end{array} \label{Jung}
\end{equation}
Therefore $q_i$ is the length of the $i$-th line while the hight of the $l$-th
column is $c_l$.

2. Let us consider the arbitrary tensor $\Psi_{i^1_1 ...i^1_{q_1} ...i^{c_1}_1
... i^{c_1}_{q_{c_1}}}$. The tensor with the required symmetry can be obtained
by means of the following transformation:
\begin{eqnarray}
\Psi_{i^1_1 ...i^1_{q_1} ...i^{c_1}_1 ... i^{c_1}_{q_{c_1}}} \rightarrow {\cal
S}^q \Psi_{i^1_1 ...i^1_{q_1} ...i^{c_1}_1 ... i^{c_1}_{q_{c_1}}} \nonumber\\
= \sum_{S} (-)^{P(S)} \Psi_{S[i^1_1 ...i^1_{q_1} ...i^{c_1}_1 ...
i^{c_1}_{q_{c_1}}]},\label{o}
\end{eqnarray}
where the sum is over the permutations $S$ of the indices. Each permutation
that enters the sum consists of the ordered product ${\cal Q} \times {\cal P}$
of some permutation $\cal P$ inside the lines $(i^k_1 ...i^k_{q_k})$ of
(\ref{Jung}) and some permutation $\cal Q$ inside the columns $(i^1_l
...i^{c_l}_{l})$. $P(S) = 0(1)$ if $Q$ is even (odd).

3. Let $U_{ij}$ be the $N\times N$ matrix that represents a certain element of
$SU(N)$. Then the action of this element on $\cal V$ is given by:
\begin{eqnarray}
\Psi_{i_1 i_2 ... i_r}  \rightarrow {\cal D}_q (U)^{j_1 j_2 ... j_r}_{i_1 i_2
... i_r} \Psi_{j_1 j_2 ... i_r} \nonumber\\
= U_{i_1 j_1} U_{i_2 j_2}... U_{i_r
j_r} \Psi_{j_1 j_2 ... j_r}\label{q}
\end{eqnarray}
We shall also write:
\begin{equation}
|\Psi\rangle \rightarrow {\cal D}_q (U) |\Psi\rangle
\end{equation}

4. Let us represent an infinitesimal element of $SU(N)$  as $ U = {\rm exp} (i
A s)$. Here $s$ is an infinitely small parameter and $A$ is a matrix that
represents a certain element of the Lie algebra $su(N)$. Then (\ref{q}) defines
the representation of $su(N)$. We denote the correspondent matrix by ${\cal
D}_q(A)$:
\begin{equation}
{\cal D}_q (U) = {\rm exp} (i {\cal D}_q(A) s)
\end{equation}

5. Using the Cartan basis any $A \in su(N)$ can be represented as
\begin{eqnarray}
A = \sum_{i = 1,...,N-1} H^i {\rm Tr}\, A H^i + \nonumber\\ \sum_{i<j; i,j =
1,...,N-1}(E^{ij} ({\rm Tr} A E^{ij})^* + (E^{ij})^+ {\rm Tr} A
E^{ij})\label{Car}
\end{eqnarray}
The Cartan elements of the basis are
\begin{equation}
 H^i = \frac{1}{\sqrt{i(i+1)}} {\rm diag}\, (1, ..., 1, -i, 0,
 ...)
\end{equation}
The shift operators are:
\begin{equation}
 (E^{ij})_{ab} = \delta_{ai}\delta_{bj}
\end{equation}

6. The normalized vector $|\Lambda^a\rangle$ ($\langle
\Lambda^a|\Lambda^a\rangle = 1$), which is an eigenvector for any operator
${\cal D}_q (H^i)$ is called the weight vector. Here $\Lambda^a \, (a = 1, ...,
N-1)$ are the correspondent eigenvalues:
\begin{equation}
{\cal D}_q (H^i) |\Lambda^a\rangle = \Lambda^i | \Lambda^a\rangle
\end{equation}

Shift operators act on the weight vectors as follows:
\begin{eqnarray}
{\cal D}_q (E^{ij})|\Lambda^a\rangle &\sim& |\Lambda^a + (\alpha^{ij})^a\rangle
\nonumber\\ {\cal D}_q (E^{ij})^+|\Lambda^a\rangle &\sim& |\Lambda^a
-(\alpha^{ij})^a\rangle
\end{eqnarray}
Here $|a\rangle \sim |b\rangle$ means that $|a\rangle = C |b\rangle$ for some
constant $C$. The numbers $\alpha^{ij}$ are called the roots of $su(N)$. They
enter the following commutation relations:
\begin{eqnarray}
&& [H^a, E^{ij}] =  (\alpha^{ij})^a E^{ij} \nonumber\\
&& [H^a,(E^{ij})^+]  =  -(\alpha^{ij})^a (E^{ij})^+
\end{eqnarray}

7. The weight vector that is annihilated by each of the operators $E^{ij}$ is
called the eldest vector. The correspondent set of numbers $\Lambda^a$ is
called the highest weight and is denoted by $m^a$.

We denote the eldest vector by $|0\rangle$. Up to the normalization factor it
corresponds to the tensor
\begin{equation}
\Psi^0_{i^1_1 ...i^1_{q_1} ...i^{c_1}_1 ... i^{c_1}_{q_{c_1}}}
 \sim  {\cal S}^q \delta^1_{i^1_1} ... \delta^1_{i^1_{q_1}}
  ...\delta^{c_1}_{i^{c_1}_1} ... \delta^{c_1}_{i^{c_1}_{q_{c_1}}}
\end{equation}

8. The highest weight of the representation $q$ is given by:
\begin{equation}
m_{i} = \frac{1}{\sqrt{i(i+1)}}(q_1 + q_2 + ... + q_{i-1} - i q_{i})
\end{equation}

9. The coherent state system correspondent to the representation $q$ is the
following subset of $\cal V$:
\begin{equation}
\{| g \rangle \, :\, | g \rangle = e^{i \phi(g)} {\cal D}_q(g) | 0
\rangle \, (g \in SU(N)) \},
\end{equation}
where $\phi(g)$ is a real - valued function on $SU(N)$.

10. The coherent state system possesses the following completeness relation:
\begin{equation}
\int d g |g \rangle \langle g | = \frac{1}{D(q)}
\end{equation}
Here $D(q)$ is the dimension of the representation. The measure on $SU(N)$ is
normalized in such a way that $\int d g = 1$.

\section{The Abelian representation for  the non -  Abelian Wilson loop on the lattice.}

In this section we consider  the lattice $SU(N)$ Wilson loop.
 As an auxiliary
instrument we use the lattice $SU(N)$ gauge model coupled to the scalar field
$\Phi$ that belongs to the irreducible representation $q$ of $SU(N)$. In
addition this scalar field is considered in the presence of the external $U(1)$
gauge field $\theta\in ]-\pi, \pi]$ (that lives on the links of the lattice).
 The partition function
has the form:
\begin{equation}
Z[\theta] = \langle \int D\Phi exp(\sum_{xy} \Phi^+_x {\cal
D}_q(U_{xy}) e^{i\theta} \Phi_y - V(\Phi))\rangle   \label{F}
\end{equation}
Here $U \in SU(N)$ and the average is calculated in a certain $SU(N)$ gauge
theory. We assume that the potential $V$ is infinitely deep and that it has the
minimum at $\Phi = |0\rangle$ with $V(|0\rangle  ) = 0$, where $|0\rangle$ is
the eldest vector of the representation. We also assume that
 $V$ is invariant
under the transformation $\Phi \rightarrow e^{i\alpha} {\cal
D}_q(g) \Phi$, where  $\alpha$ is arbitrary real-valued field.

The idea is that there exist two different ways to rewrite the partition
function as a sum over the worldlines of the scalar. The first one demonstrates
that the partition function of the model generates the non - Abelian Wilson
loop average (in the representation $q$). The second one demonstrates that the
same functional also generates the Abelian - like expression for the Wilson
loop. Finally due to the arbitrariness of the $SU(N)$ measure we obtain the
equivalence of the Wilson loop itself and its Abelian - like representation.

\subsection{$Z[\theta]$ generates the Wilson loops}

 The
mentioned properties of the potential make it possible to reduce the
integration over $\Phi$ to the integration over  $g \in SU(N)$ and over a real
- valued scalar field $\alpha$. Both fields are defined on the sites of the
lattice. Then we can rewrite (\ref{F})
 as a sum over the
worldlines of the scalar as follows:
\begin{eqnarray}
&& Z[\theta]  = \langle \int D\alpha Dg exp( \sum_{xy} \langle 0|{\cal
D}_q(g)^+_x {\cal D}_q(U_{xy})\nonumber\\&&
e^{i\theta_{xy} + i\alpha_y - i\alpha_x}{\cal D}_q(g)_y|0\rangle   )\rangle  \nonumber\\
& = &  \langle \int Dg \sum_{\cal C} (2\pi)^{n(j({\cal C}))} e^{i(j({\cal
C}),\theta)} \nonumber\\&&{\rm  Tr}\Pi_{\{xy\}\in {\cal C}} \frac{1}{|j({\cal
C})_{xy}|!} \langle g_x|{\cal D}_q(U_{xy})|g_y\rangle ^{|j({\cal C})_{xy}|}
\rangle \label{Z1}
\end{eqnarray}

Here we used the calculus of differential forms on the lattice (for the
definitions see, for example, \cite{forms}). The sum is over the closed lattice
paths $\cal C$. The closed lattice path is defined as a map ${\cal C}: Z_N =
\{0, 1, ..., N-1\} \rightarrow X$ (where $X$ is a set of the lattice points and
$N$ is an integer number) such that for each $k \in Z_N$ the points ${\cal
C}(k)$ and ${\cal C}({[k+1]{\rm mod}\,N})$ are connected by a link. The product
over all such links is denoted by $\Pi_{\{xy\}\in {\cal C}}$, where each link
is counted once and it is implied that $x = {\cal C}(k)$ and $ y = {\cal
C}([k+1]{\rm mod}\, N)$ for some $k$. $j({\cal C})$ is an integer - valued link
field that corresponds to the path ${\cal C}$. $|j_{xy}|$ counts  the number of
times the given link $\{xy\}$ is encountered when we are moving along the path.
The sign of $j_{xy}$ reflects the orientation of the path $\cal C$. So the case
$|j|
> 1$ corresponds to the overlapping of the different pieces of $\cal C$. It is
worth mentioning that $j$ does not contain the complete information on the
orientation of $\cal C$. Therefore different paths may correspond to the same
$1$ - form $j({\cal C})$. This may happen if ${\cal C}$ has the
selfintersections. However, in (\ref{Z1}) the sum is over $\cal C$  that have
{\it different} $j({\cal C})$. This means that the sum is indeed over different
$j({\cal C})$ and each term of the sum depends only upon $j({\cal C})$ (not on
$\cal C$ itself). $n(j)$ is the number of points, which enter the path
represented by $j$. Also we assume the following notation:
\begin{equation}
(j,\theta) = \sum_{xy} j_{xy} \theta_{xy},
\end{equation}
where the sum is over the links of the lattice.

We denote $\langle i|{\cal D}_q(g)|0\rangle   = {\cal D}_q(g)_{i0};\langle
0|{\cal D}_q(g)^+|i\rangle   = {\cal D}_q(g)^+_{0i}$, where $|i\rangle  $ is
the basis vector of the  space $\cal V$ of the representation $q$. The
integration over $Dg$ can be performed using the following expressions:
\begin{eqnarray}
&& \int dg {\cal D}_q(g)^+_{i0}{\cal D}_q(g)_{0j} = \frac{\bf 1}{M_1(q)}\delta_{ij}\nonumber\\
&& \int dg {\cal D}_q(g)^+_{i_10}{\cal D}_q(g)_{0j_1}{\cal D}_q(g)^+_{i_20}{\cal D}_q(g)_{0j_2} = \nonumber\\
&&\frac{\bf 1}{M_2(q)}(\delta_{i_1j_1}\delta_{i_2j_2}+\delta_{i_1j_2}\delta_{i_2j_1})\nonumber\\
&& ... \label{Int}
\end{eqnarray}
Here the measure $dg$ is normalized in such a way that $\int dg = 1$. The first
identity reflects the completeness of a coherent state system and implies that
$M_1(q) = D(q)$, where $D(q)$ is the dimension of the representation. Other
equations follow from the gauge invariance. The correspondent coefficients
$M_i(q)$ depend on the representation and can be calculated explicitly using
the expressions for the basis vectors of $\cal V$ and the technique of a strong
coupling expansion (see, for example, \cite{Creutz}).

We obtain:
\begin{equation}
 Z[\theta]  =  \langle \sum_{{\cal C}}C(j({\cal C}))
e^{i(j({\cal C}),\theta)} W_q[{\cal C}] \rangle \label{W1}
\end{equation}
where the sum is over {\it all different} $\cal C$. $W_q[{\cal C}]$ is the
Wilson loop (in the representation $q$) correspondent to the path ${\cal C}:
Z_N \rightarrow X$:
\begin{eqnarray}
W_q[{\cal C}]& = &{\rm Tr} {\cal D}_q(U_{{\cal C}(0) {\cal C}(1)}){\cal
D}_q(U_{{\cal C}(1) {\cal C}(2)})...\nonumber\\&&...{\cal D}_q(U_{{\cal C}(N-1)
{\cal C}(0)})
\end{eqnarray}
We also denote
\begin{equation}
 C(j) =
\frac{(2\pi)^{n(j({\cal C}))}}{(M_1(q))^{n_1(j)}(M_2(q))^{n_2(j)}...}
\label{C1}
\end{equation}
Here $n_1$ is the number of points  of the path ${\cal C}$, where
 there are no selfintersections, $n_2$ is the number of
points where ${\cal C}$ intersects itself once (i.e. where there are two
different $k_a \in Z_N, (a = 1,2)$, such that ${\cal C}(k_1) = {\cal C}(k_2)$),
$n_2$ is the number of points where ${\cal C}$ intersects itself twice  etc ($n
= \sum_i n_i$).

As we have already mentioned, $j$ does not contain the complete information on
the orientation of $\cal C$. Hence we extract from (\ref{W1}) the sum of the
Wilson loops, which have the same $j({\cal C})$. The expression for the vacuum
average of $W_q^{l} = \sum_{j[{\cal C}] = l}
 W_q[{\cal C}]  $ can be obtained using the Fourier transformation:
\begin{eqnarray}
\langle W_q^l\rangle   = \sum_{j[{\cal C}] = l} \langle W_q[{\cal C}]\rangle
\nonumber\\ = \frac{1}{C[{l}]}\int D \theta exp(- i(l, \theta)) Z[\theta]
\end{eqnarray}

\subsection{$Z[\theta]$ generates the Abelian - like expressions for the Wilson
loop}

Let us fix the gauge, where $\Phi_x = e^{i\alpha_x}|0\rangle  $ ($\alpha$ is a
real valued field). We get:

\begin{eqnarray}
&&Z[\theta] = \langle \int D\alpha \,{\rm exp}( \sum_{xy} \langle 0|{\cal
D}_q(U_{xy})|0\rangle   e^{i(\theta_{xy} + \alpha_x - \alpha_y)}) \nonumber\\
&& = \langle \sum_{\cal C} \bar{C}(j({\cal C}))
 \,{\rm  exp}(i (j({\cal C}),\theta)) \nonumber\\&&
 \Pi_{\{xy\}\in{\cal C}}\frac{1}{|j({\cal C})_{xy}|!}\langle 0|{\cal D}_q(U_{xy})
 |0\rangle ^{|j({\cal C})_{xy}|}  \rangle \label{W2}
\end{eqnarray}

Here the sum is over $\cal C$ that have different $j({\cal C})$ and
\begin{equation}
\bar{C}(j) = (2\pi)^{n(j)}
\end{equation}

\subsection{Equivalence of the Wilson loop and its Abelian representation}
 Recall that (\ref{W1}) and (\ref{W2}) are the different
 representations of the same quantity. Hence we obtain:
\begin{eqnarray}
&&\langle W_q^l\rangle   = \frac{1}{C(l)}\int D \theta \,{\rm exp}\,(-i (l,
\theta)) Z[\theta]\nonumber\\
&& =  \sum_{j[{\cal C}] = l}\frac{\bar{C}(l)}{C(l)}\langle \Pi_{\{xy\}\in {\cal
C}}\frac{1}{|j({\cal C})_{xy}|!}\langle 0|{\cal D}_q(U_{xy})|0\rangle
^{|j({\cal C})_{xy}|}  \rangle \nonumber\\
&&= \sum_{j[{\cal C}] = l}\tilde{C}(l)\langle \int Dg \Pi_{\{xy\}\in {\cal C}}
\langle 0|{\cal D}_q(U^g_{xy})|0\rangle ^{|j({\cal C})_{xy}|} \rangle ,
\label{Rp}
\end{eqnarray}
where we used the gauge invariance of the operation $\langle  ... \rangle $ and
introduced a new combinatorial factor
\begin{equation}
\tilde{C}(l) = (M_1(q)^{n_1(l)}M_2(q)^{n_2(l)}...)\Pi_{\rm
links}\frac{1}{|l_{\rm link}|!},
\end{equation}
where the sum is over all links of the lattice. The gauge transformation of the
link variable $U_{xy}$ is denoted by $U_{xy}^g = g U_{xy} g^+$. The measure $D
g$ is defined in such a way that $\int D g = 1$.

Due to the arbitrariness of the  averaging $\langle ... \rangle$ over the gauge
field $U$ we have:
\begin{equation}
W_q^l = \sum_{j[{\cal C}] = l}\tilde{C}[l] \int D g \Pi_{\{xy\}\in {\cal
C}}\langle 0|{\cal D}_q(U^g_{xy})|0\rangle ^{|l_{xy}|} \label{fin1}
\end{equation}

Thus the sum of $SU(N)$ Wilson loops that are correspondent to the same $1$ -
form $l$ can be represented through the Abelian - like expressions
$\Pi_{\{xy\}\in {\cal C}}\langle 0|{\cal D}_q(U^g_{xy})|0\rangle ^{|l_{xy}|}$.
For the Wilson loops without selfintersections we have:
\begin{equation}
W^{l[{\cal C}]} = W[\cal C]
\end{equation}
Then $\tilde{C}[l] = D(q)^{n(l)} = D(q)^{|{\cal C}|}$, where $|{\cal C}|$ is
the length of the loop. In this case (\ref{fin1}) acquires the form:
\begin{equation}
W_q[{\cal C}] = D(q)^{|{\cal C}|} \int D g \Pi_{{xy}\in {\cal
C}}\langle 0|{\cal D}_q(U^g_{xy})|0\rangle ,\label{fin2}
\end{equation}

The expression (\ref{fin2}) can be considered as the main result of this paper.
In the next section we shall obtain the explicit expression for the matrix
elements entering (\ref{fin2}) and discuss symmetry properties of (\ref{fin2}).

\section{Properties of the derived representation.}

In this section we derive the expression for $\langle 0|{\cal
D}_q(U^g_{xy})|0\rangle$ entering (\ref{fin2}). Also we introduce some new
notations in order to demonstrate the Abelian nature of (\ref{fin2}).

We use the explicit expressions for the eldest vector and for the matrix ${\cal
D}_q(U)$ given in the section $2$. We can represent $\langle 0|{\cal
D}_q(U)|0\rangle$ as
\begin{equation}
\langle 0|{\cal D}_q(U)|0\rangle = {\cal U}(U),
\end{equation}
where
\begin{eqnarray}
{\cal U}(U) = \frac{1}{K}\sum_{S,\tilde{S}}
(-)^{P(S)+P(\tilde{S})}
 U_{\tilde{S}^1_1 S^1_1} ... U_{\tilde{S}^1_{q_1} S^1_{q_1}}\nonumber\\
 ... U_{\tilde{S}^{c_1}_1 S^{c_1}_1}...U_{\tilde{S}^{c_1}_{q_{c_1}}
 S^{c_1}_{q_{c_1}}}\label{Up}
\end{eqnarray}
and the normalization factor $K$ is chosen in such a way that ${\cal U}(1) =
1$. Here the sum is over $S^j_{i}$ and $\bar{S}^j_{i}$  that are the elements
of the following tables:
\begin{equation}
\begin{array}{cc}
\begin{array}{ccccc}
S^1_1 & ... & S^1_i & ... & S^1_{q_1} \\
S^2_1 & ... & S^2_{q_2} \\
... \\
S^{c_1}_1 & ...
\end{array}\,\,\, & \,\,\,
\begin{array}{ccccc}
\bar{S}^1_1 & ... & \bar{S}^1_i & ... & \bar{S}^1_{q_1} \\
\bar{S}^2_1 & ... & \bar{S}^2_{q_2} \\
... \\
\bar{S}^{c_1}_1 & ...
\end{array}
\end{array}
\end{equation}
(The lines contain $q_1 , q_2, ..., q_{c_1}$ elements while the columns contain
$c_1 , c_2, ..., c_{q_1}$ elements respectively.) Each of these two  tables is
obtained via the composition of some permutation inside the lines and some
permutation inside the columns of the following table:
\begin{equation}
\begin{array}{ccccc}
1 & ... & 1 & ... & 1 \\
2 & ... & 2 \\
... \\
c_1 & ...
\end{array}
\end{equation}
$P(S) = 0(1)$ if the correspondent column permutation is even (odd).

We can rewrite (\ref{Up}) in a compact form
\begin{equation}
{\cal U}(U) = \omega_{c_1}(U) \omega_{c_2}(U) ... \omega_{c_{q_1}}(U),\label{U}
\end{equation}
where
\begin{equation}
\omega_M (U) = {\rm det} \left(
\begin{array}{ccc}
 U_{1 1} & ...& U_{1 M}\\
  ... & ... & ... \\
 U_{M 1} & ...& U_{M M}
\end{array}\right)\label{U}
\end{equation}

In order to demonstrate the $U(1)$ nature of the derived expression for the
Wilson loop (\ref{fin2}) we rewrite it as follows:
\begin{equation}
W_q[{\cal C}] =  \int D g {\rm exp} \,(i({\cal A}^g,j({\cal C})) +
{\rm log}\, (D(q)) |{\cal C}|), \label{fin3}
\end{equation}
where
\begin{equation}
{\cal A}^g_{xy} = - i {\rm log}\, {\cal U}(U^g_{xy})
\end{equation}
is a complex - valued link field. Below we shall show that its real part can be
treated as a lattice Abelian gauge field. (It is worth mentioning that its
imaginary part vanishes in a continuum limit. This will be shown in the next
section, when we shall consider the continuum limit of the derived
expressions.)

The exponent in (\ref{fin3}) is invariant under the $U(N-c_1)$ - transformation
\begin{equation}
U_{xy} \rightarrow h^+(\Omega_x)U_{xy}h(\Omega_y)\,
\end{equation}
where $\Omega_x \in U(N - c_1)$ is attached to the points of the path $\cal C$,
\begin{equation}
h(\Omega) = \left( \begin{array}{ccccc}
e^{i\alpha} &  0           & ... & ...            & ... \\
0           &  e^{i\alpha} & ... & ...            & ... \\
...         &  ...         & ... & ...            & ... \\
...         &  ...         & 0   & e^{i\alpha} & 0   \\
...         &  ...         & ... & 0              & \Omega
\end{array} \right),
\end{equation}
and
\begin{equation}
{\rm  det}\Omega = e^{-i c_1 \alpha}
\end{equation}

This transformation acts on ${\cal A}^g$ as follows:
\begin{equation}
{\cal A}_{xy}^g \rightarrow {\cal A}_{xy}^{hg} = {\cal A}_{xy}^g + r (\alpha_x
- \alpha_y),
\end{equation}
where $r = \sum_i q^i$ is the rank of the representation. This demonstrates
that the mentioned $U(N-c_1)$ symmetry induces the $U(1)$ gauge symmetry and
real part of ${\cal A}^g$ plays the role of the correspondent gauge field.

In analogy with the continuum theory we obtain the lattice variant of the non -
Abelian Stokes theorem:
\begin{equation}
W_q[{\cal C}] =  \int D g {\rm exp} \,(i(d {\cal A}^g,m({\cal C}))
+ {\rm log}\, (D(q)) |{\cal C}|), \label{fin4}
\end{equation}
where $m({\cal C})$ is the integer - valued plaquette variable that represents
the surface that spans the contour $\cal C$. $d {\cal A}^g$ is the plaquette
variable that is defined in an analogy with the Abelian field strength:
\begin{equation}
[d {\cal A}]_{xyzw} = {\cal A}_{xy} + {\cal A}_{yz} - {\cal A}_{wz} - {\cal
A}_{xw},
\end{equation}
where we suppose that the directions $(xy)$ and $(yz)$ are positive. The sum
over the plaquettes is denoted by
\begin{equation}
(m,d{\cal A}) = \sum_{xyzw} m_{xyzw} [d{\cal A}]_{xyzw}
\end{equation}

Let us mention two particular cases, when expression (\ref{U}) has the most
simple form.

For the symmetric ($q_1 = r, q_2 = ... = q_{N-1} = 0$) representations we have:
${\cal U} = U_{11}^r$. Thus the expression for the Wilson loop has the form:
\begin{equation}
W_r[{\cal C}] =  \int D g {\rm exp} \,(r({\rm log}U^g_{11} ,j({\cal C})) + {\rm
log}\, (D(q)) |{\cal C}|)
\end{equation}
(The adjoint representation is an example of such a representation. It
corresponds to $r = 2$.)

Another simple case corresponds to the absolutely antisymmetric representation
($q_1 = q_2 = ... = q_{r} = 1, q_{r+1} = ... = q_{N-1} = 0$). In this case we
have:
\begin{equation}
{\cal U} = {\rm det} \left( \begin{array}{ccc}
 U_{1 1} & ...& U_{1 r}\\
  ... & ... & ... \\
 U_{r 1} & ...& U_{r r}
\end{array}\right)
\end{equation}

\section{The continuum limit.}

In order to consider consistently the continuum limit of a lattice model, its
behavior in a small vicinity of the critical couplings should be investigated.
Only near the point of the second order phase transition the lattice
correlation length becomes infinite and the lattice spacing can be considered
as infinitely small. In the correspondent model the continuum gauge field that
is not smooth may arise. It was shown, for example, that such singular
configurations of a gauge field as the center vortices may survive a continuum
limit \cite{BVZ}.

In this paper we consider only smooth continuum gauge fields. The continuum
limit of the derived lattice expressions in the case of singular field
configurations is out of scope of the present consideration.

Let us consider the expression (\ref{fin3}) within   the lattice $SU(N)$ gauge
model that has the second order phase transition. Let us also consider the
behavior of the following correlator:
\begin{equation}
<O(U)W_q[{\cal C}]>
\end{equation}
Here $O(U)$ is gauge invariant functional. We assume that $\cal C$ has no
selfintersections. We have:

\begin{eqnarray}
\langle O(U)W_q[{\cal C}]\rangle = \langle O(U) {\rm exp} \,(i({\cal A},j({\cal
C}))\nonumber\\ + {\rm log}\, (D(q)) |{\cal C}|)\rangle ,\label{Av}
\end{eqnarray}
Here the integration over gauge transformations was absorbed by the average
$\langle ...\rangle$. Let us suppose that the singular gauge fields do not
arise in the continuum limit of the considered model. Then we can use the
correspondence:
\begin{equation}
U_{xy} \rightarrow (1 + i A_{\mu} \Delta x), \label{UA}
\end{equation}
where $A$ is a continuum gauge field and $\Delta x$ is an infinite small space
distance. In accordance with (\ref{UA}) $\langle 0|{\cal D}_q(U)|0\rangle$
corresponds to
\begin{equation}
\langle 0|{\cal D}_q(U_{xy})|0\rangle   \rightarrow 1 + i \langle 0| {\cal
D}_q(A_{\mu})|0\rangle \Delta x
\end{equation}
Hence it is clear that
\begin{equation}
{\cal A}_{xy}\rightarrow \langle 0| {\cal D}_q(A_{\mu})|0\rangle \Delta x
\end{equation}
and imaginary part of  ${\cal A}$  vanishes in a continuum limit.

Recall that in the vicinity of the second order phase transition ${\cal C} =
{\cal C}_{\rm phys} \frac{1}{a}$, where the lattice spacing $a$ tends to zero
and the physical length of the contour is equal to ${\cal C}_{\rm phys} =
\int_{\cal C} \sqrt{dx_{\mu}dx_{\mu}}$. Let us consider such $O$ that it tends
to a smooth functional of $A$ in the continuum limit. Therefore we can write
the continuum limit of (\ref{Av}) as follows:
\begin{eqnarray}
\langle O(U)W_q[{\cal C}]\rangle = \langle O(A) {\rm exp} \,(i \int \langle 0|
{\cal D}_q(A_{\mu}) | 0 \rangle \, d x_{\mu} \, \nonumber\\+ \, m_q \int_{\cal
C} \sqrt{dx_{\mu}dx_{\mu}}) \rangle
\end{eqnarray}
where the constant $m_q$ is divergent as
\begin{equation}
m_q  \sim {\bf M} {\rm log} D(q),\label{Mq}
\end{equation}
Here ${\bf M}$ is the ultraviolet cut-off (which in the case of the lattice
regularization is equal to the inverse lattice spacing).

In gauge theory any Wilson loop always corresponds to the charged particle. The
divergence of $m_q$ in (\ref{Mu}) should be included into the renormalization
scheme for the mass of this particle.

Let us restore the integration over the gauge transformations. The obtained
continuum theory is gauge invariant. Hence we can rewrite the last expression
as:
\begin{eqnarray}
\langle O(U)W_q[{\cal C}]\rangle &=&  \langle O(A) \int D g {\rm exp} \,( i
\int \langle 0| {\cal D}_q(A_{\mu}) | 0 \rangle \, d x_{\mu} \, \nonumber\\&& +
\, m_q \int_{\cal C} \sqrt{dx_{\mu}dx_{\mu}}) \rangle,
\end{eqnarray}
where $D g $ is the measure on the space of all $\it smooth$ gauge
transformations, defined in such a way that $\int D g = 1$.

Due to the arbitrariness of the average $ \langle ... \rangle$ we have:
\begin{equation}
W_q[{\cal C}] = \int D \mu_{\cal C}(g) {\rm exp} \,(i \int \langle
0| {\cal D}_q(A_{\mu}) | 0 \rangle \, d x_{\mu} ),\label{final}
\end{equation}
where
\begin{equation}
D \mu_{{\cal C},q}(g) = {\cal F}({\cal C}) Dg = {\rm exp} ( m_q \int_{\cal C}
\sqrt{dx_{\mu}dx_{\mu}}) Dg \label{Mu}
\end{equation}
This measure differs from the conventional one, which is normalized in such a
way that $\int Dg = 1$. Formally ${\cal F} \rightarrow \infty$ in the continuum
limit. Thus if we substitute $Dg$ into (\ref{Wdp}) instead of $D \mu_{{\cal
C},q}(g)$, then the integral vanishes identically in the continuum limit. (This
vanishing was pointed out in \cite{Iv} for some particular choices of $A$.)

Of course, in any further application of this formula we should remember that
the integration is over the smooth gauge transformations and the formula itself
is valid only within the model, in which the singular gauge configurations are
forbidden.

We use Cartan representation (\ref{Car}) of $A^g$  and obtain
\begin{eqnarray}
\langle 0|{\cal D}_q(A^g)|0\rangle   =  \langle 0| \sum_{i = 1,...,N-1} H^i
{\rm Tr}\, A^g H^i |0\rangle \nonumber\\ =   {\rm Tr}\, A^g {\cal H}^q
\end{eqnarray}

Thus the representation (\ref{final}) has the form (\ref{Wdp}) originally
proposed by D.Djakonov and V.Petrov \cite{DP}.

Using the expressions for the Cartan elements $H^i$ and for the highest weight
$m^i$ it is easy to calculate  ${\cal H}^q $. It's nonzero elements are:
\begin{equation}
 {\cal H}^{q}_{ii} = q_i - \frac{1}{N} \sum_{k = 1, ..., N} q_k \, (i = 1, ..., N),\label{H}
\end{equation}
where we imply that $q_N = 0$.

Hence, we can rewrite (\ref{final}) as follows:
\begin{equation}
W_q[{\cal C}] = \int D \mu_{\cal C}(g) {\rm exp} \,(i \int \sum_{i = 1, ...,
N-1} ([A^g_{\mu}]_{ii} q_i) d x_{\mu} )\label{final1}
\end{equation}

It is worth mentioning that this expression can be obtained directly from
(\ref{fin3}) (with $\cal U$ that is given in (\ref{U})).

\section{Conclusions}

In this work we have derived the  Abelian representation (\ref{fin3}) for the
$SU(N)$ Wilson loop on the lattice. This representation contains the complex -
valued link field. This field is a function of the matrix elements of the
$SU(N)$ gauge field. (\ref{fin3}) possesses the $U(1)$ gauge symmetry. Real
part of the mentioned link field can be considered as a correspondent gauge
field. Its imaginary part vanishes in a continuum limit.

The Non - Abelian Stokes theorem on the lattice commonly follows from
(\ref{fin3}). It can be used for "counting"  of the "magnetic" flux that is
contained inside the contour correspondent to the Wilson loop. Let us notice
that both in the $P$ - vortex and in the monopole pictures of a confinement
phenomenon the Wilson loop is approximated by certain Abelian Wilson loop. The
Abelian magnetic flux inside the correspondent contour appears to be
responsible for the confinement. Therefore "counting" of the "magnetic" flux
performed with the help of the non - Abelian Stokes theorem can be used in
related confinement pictures in order to avoid an inexactitude associated with
the mentioned approximation.

It is worth mentioning that the continuum limit of the derived expression
coincides with the original representation obtained by D. Diakonov and V.
Petrov.

\begin{acknowledgments}
The author is grateful to A.N.Ivanov for the discussions, which stimulated him
to investigate the subject considered in this paper. This work was partly
supported by the grants INTAS 00-00111 and CRDF award RP1-2364-MO-02.
\end{acknowledgments}

\end{document}